\documentclass[12pt,prd,draft,nofootinbib,floatfix]{revtex4}

\usepackage{amssymb}
\usepackage{amsmath}
\usepackage{bm}
\usepackage{latexsym}
\newcommand{\beq}{\begin{eqnarray}}
\newcommand{\eeq}{\end{eqnarray}}
\newcommand{\be}{\begin{equation}}
\newcommand{\ee}{\end{equation}}
\def\la{\mathrel{\mathpalette\fun <}}
\def\ga{\mathrel{\mathpalette\fun >}}
\def\fun#1#2{\lower3.6pt\vbox{\baselineskip0pt\lineskip.9pt
\ialign{$\mathsurround=0pt#1\hfil ##\hfil$\crcr#2\crcr\sim\crcr}}}
\newcommand{{\SD}}{\rm SD}

\newcommand{\vep}{\bm p}

\begin{document}

\title{The charmed mesons in the region above 3.0~GeV}
\author{\firstname{A.M.}~\surname{Badalian}}
\email{badalian@itep.ru} \affiliation{Institute of Theoretical and
Experimental Physics, Moscow, Russia}
\author{\firstname{B.L.G.}~\surname{Bakker}}
\email{b.l.g.bakker@vu.nl} \affiliation{Department of Physics
and Astronomy, Vrije Universiteit, Amsterdam, The Netherlands}

\date{\today}

\begin{abstract}
The masses of excited charmed mesons are shown to decrease by $\sim (50-150)$~MeV due to a flattening of the confining potential at large distances, which
effectively takes into account open decay channels. The scale of the mass shifts  is similar to that in charmonium for $\psi(4660)$ and $\chi_{c0}(4700)$.
The following masses of the first excitations: $M(2\,{}^3P_0)=2874$~MeV, $M(2\,{}^3P_2)=2968$~MeV,  $M(2\,{}^3D_1)=3175$~MeV, and $M(2\,{}^3D_3)=3187$~MeV,
and second excitations: $M(3\,{}^1S_0)=3008$~MeV,  $M(3\,{}^3S_1)=3062$~MeV,  $M(3\,{}^3P_0)=3229$~MeV, and $M(3\,{}^3P_2) =3264$~MeV, are predicted.  The
other states with $L=0,1,2$ and  $n_r \geq 3$ have their masses in the region $M(nL)\geq 3.3$~GeV. 

\end{abstract}

\maketitle

\section{Introduction}
\label{Sec.I}
Recently new charmed mesons in the region (3.0 - 3.2)~GeV were observed \cite{1,2,3,4,5} and theoretical studies of their quantum numbers,
hadronic decays, and other properties were presented \cite{6,7,8,9,10,11,12,13}. The new structures, $D_2^*(3000)^0$ with $J^P=2^+$, $D(3000)^0$
and $D_J^*(3000)^0$, have rather large  excitation energies (above the ground states), $E\sim 1$~GeV, which, however, remain significantly smaller
than those in the charmonium family, where the maximal excitation energy  $E_{\rm max}(c\bar c)= M(\psi(4660)) - M(J/\psi) = 1.546(9)$~GeV and in light mesons
$M(\rho(2150)) - M(\rho(1S))\sim 1.40(5)$~GeV. Therefore, according to modern representations about the structure of conventional mesons, one can expect
of charmed resonances to exist in the range up to 3.6 GeV and theoretical studies of the $c\bar q$ spectrum were already presented in different approaches:
in  relativistic potential models \cite{6,10,13,14}, the relativistic string model \cite{15,16}, and other approaches \cite{9,11,12}. However, the predicted masses
of high excitations, even within the same relativistic Hamiltonian, strongly depend on the parameters used \cite{6,13,14}, in particular, on the value of
the string tension of the confining potential (CP). To better  understand  the differences in the predictions it is useful to compare only relativistic
models \cite{6,14,15}, where the same string tension $\sigma=0.180$~GeV$^2$ is used. This correct choice of $\sigma$ is specifically important for high excitations,
which masses are mostly determined by the CP, while the gluon-exchange (GE) potential, even with large strong coupling, contributes $\la 10\%$ to the mass.

In a strict sense, the masses of high resonances have to be determined solving a many-channel system, as in the nonrelativistic model for heavy quarkonia
(HQ) \cite{17,18},  or in the relativistic many-channel model, developed in Refs.~\cite{19,20}. In the relativistic case, this task is very complicated,
since high excitations have a large number of important decay  channels, e.g. in the $\,^3P_0$ model \cite{6} the $D_2^*(3\,^3P_2)$ resonance (with the mass 3353 MeV and
the total width 114 MeV) has eleven decay channels with almost equal branching ratios $\sim (3-17)\%$. However, the mass shifts due to decay channels cannot be defined in the
$\,^3P_0$ model.  Instead, as in light mesons, the hadronic shifts can be calculated in a single-channel approximation, if the flattening of the  CP,  $V_{\rm fl}(r)$,
at large distances, due to influence of open channels, is taken into account. This potential was introduced long ago while performing  an analysis of the radial excitations of light mesons
\cite{21,22} and later, assuming the universal
character of the flattened potential, a good description of high excitations of heavy quarkonia (HQ) was obtained \cite{16,23,24}. Note that without the flattening effect
it is  impossible to get the correct slope of the radial Regge trajectories (RT)
in light mesons.  Another behavior of the CP  at large distances  was proposed in Ref.~\cite{25} and used to describe both light mesons and HQ,
while the so-called screened CP was suggested in Ref.~\cite{26} to study the HQ spectrum \cite{27} and heavy-light mesons \cite{7}.
Here we note that there exist significant differences between the flattened and the screened CP, in particular, with the screened potential
the mass shifts of high charmonium states occur to be much larger than in the flattened potential. For instance,   in Ref.~\cite{27}  the state $\chi_{c1}(3P)$
is identified  with  $X(4140)$,  while in Ref.~\cite{16}, where the flattened CP was used, the mass shift of the $3\,^3P_1$ charmonium state is
found to be  $\sim 100$~MeV smaller and this state was identified with  $X(4274)$.

The main goal of our paper is to determine the mass shifts of high charmed mesons. For that we use the relativistic string Hamiltonian (RSH)
with the universal flattened CP, applied before to light mesons and charmonium. We expect that, as in the case of charmonium,  with the flattened CP the mass shifts are not large for
excitations like  the $2P$, $2D$, and $3S$ states, but can reach $(100-150)$ MeV for higher states. We shall show that the most important factor
which determines the value  of the mass shift, is related to the size of the string, $d(nL)=\langle r \rangle _{nL}$, and for the excitations with
$d\sim (1.5-1.9)$~fm it can reach $(100-150)$ MeV. Therefore, the masses of highly excited  charmed mesons are expected to be smaller than those predicted in Refs.~\cite{6,14}.

\section{The  {\em D}-meson spectrum for the linear + gluon-exchange potential}
\label{Sec.II}

First, we present  the $D$-meson spectrum, calculated in the relativistic string Hamiltonian \cite{15,21,28}, with the light quark mass $m_q=0$,
and compare it with the spectrum obtained in two relativistic potential models which use the constituent quark masses \cite{6,14}. In all three  models
the linear term $V_{\rm c}(r)=\sigma r$ of the potential $V_0(r)$ has the  same string tension,  $\sigma=0.18$~GeV$^2$, just as in the Regge trajectories of light mesons.

In the simplified version of the RSH,
\be
H_0 = T + V_0(r),
\label{eq.01}
\ee
the spin-dependent interactions and the string corrections are considered as a perturbation, while  $V_0(r) = V_{\rm c} (r) + V_{\rm GE}(r)$
is an instantaneous potential. The kinetic term $\rm T$,
\begin{equation}
 T=\frac{\omega_q}{2}+\frac{m_q^2}{2\omega_q}+\frac{\omega_Q}{2}+  \frac{m_Q^2}{2\omega_Q}+\frac{\vep^2}{2\omega_{\rm red}},
\label{eq.02}
\end{equation}
is expressed via  the variables (the operators) $\omega_i$. Their values can be determined from the extremum conditions:
 $\frac{\partial H_0}{\partial \omega_i}=0~(i=1,2)$. It gives
\begin{equation}
\omega_i(nL)=\langle\sqrt{ \vep^2+m_i^2}\rangle_{nL} \quad (i=1,2),
\label{eq.03}
\end{equation}
i.e., $\omega_i$ is the kinetic energy operator of a quark $q_i$ and its m.e. is denoted below as $\omega_q(nL)$ for a light quark and  $\omega_c(nL)$
for the $c$-quark. The quantity $\omega_{\rm red}=\frac{\omega_q\omega_c}{\omega_q+\omega_c}$ is the reduced mass. Substituting $\omega_i$ into
Eq.~(\ref{eq.02}), one arrives at the well-known ``square-root" form of the kinetic term, denoted as $T_{\rm R}$:
\begin{equation}
 T_R=\sqrt{\vep^2+m_q^2} + \sqrt{\vep^2+m_c^2} = \sqrt{\vep^2} + \sqrt{\vep^2 + m_c^2},
\label{eq.04}
\end{equation}
where $m_q=0$ and $m_c$ is equal to the pole $c$-quark mass, $m_c=1.435$~GeV. In Ref.~\cite{6} 
the constituent quark mass, $m_q=220$~MeV, was taken in the kinetic term. The e.v. $M_0(nL)$ of  the spinless Salpeter equation (SSE),
\begin{equation}
 \left( T_{\rm R} + V_0(r)\right) \varphi_{nL}=M_0(nL)\varphi_{nL},
\label{eq.05}
\end{equation}
determines the main contribution to the spin-averaged mass  $M_{\rm cog}(nL)$, which also includes two negative corrections: the string correction
$\delta_{\rm str}(nL)$ \cite{22,24} and the  nonperturbative self-energy (SE) correction $\delta_{\rm se}(nL)$ \cite{29},
\begin{equation}
 M_{\rm cog}(nL)=M_0(nL) + \delta_{\rm str}(nL)+ \delta_{\rm se}(nL).
\label{eq.06}
\end{equation}
The masses of the $nS$ and $nL~ (L\not=0)$  states,
\be
M(nS) = M_{\rm cog} + \delta_{hf}(nS);  \quad M(nJ) = M_{\rm cog}(nL) + \delta_{\rm fs}(nJ),
\label{eq.07}
\ee
also include the hyperfine  $\delta_{\rm hf}(nS)$ or the fine-structure $\delta_{\rm fs}(nL)~(L\not= 0)$ corrections.
Notice that the self-energy (SE) correction
\be
 \delta_{\rm se}=-\frac{\sigma \eta_q}{2\omega_q(nL)} - \frac{\sigma \eta_Q}{2\omega_Q},
\label{eq.08}
\ee
is flavor-dependent: the factor $\eta_f=0.90$ for a light quark and $\eta_c=0.20$ for the $c$-quark. Due to the small value of $\eta_c$ and large
$\omega_c$ the SE contribution from the $c$-quark to $M_{\rm cog}$ is small, $\sim 10$~MeV, and later will be neglected.

The string correction \cite{21,22}
\be
\delta_{\rm str} (nl) = - \frac{L(L+1) \sigma \langle  r^{-1} \rangle_{nL}}{\omega_q ( 6 \omega_q + \langle \sigma r \rangle_{nL})},
\label{eq.09}
\ee
is expressed via the m.e. $\omega_q(nL)$. The hyperfine and the fine-structure splittings are also expressed via the kinetic energies \cite{30}
(see the Appendix). Note that in the RSH there is no a negative fitting constant $C_0$, which is present in
the constituent quark models \cite{6,14}; the presence of $C_0$ violates the linear behavior of the RT, while in the RSH the centroid mass
Eq.~(\ref{eq.06}) includes  the self-energy and the string corrections, which just provide the linear behavior of the  RT \cite{22}. It is worth to underline
that the self-energy correction decreases as a function of the orbital angular momentum $L$, while the string correction increases.

\begin{table}
\caption{The masses of low charmed mesons (in MeV) ($n_r=0,1$). Experimental data are taken from Refs.~\cite{1,2,3,4,5}}
\label{tab.01}
\begin{tabular}{|l|r|r|r|r|r|r}
\hline
State         &  GI \cite{6} &  EFG \cite{15} & this paper &  $d(nL)$ in fm  &  exp. \cite{5}\\
\hline
   $1\,^1S_0$      & 1877        & 1871         &  1869   & 0.47  &  1869.5    \\

   $1\,^3S_1$      & 2041         & 2010         &  2005   &   0.47  & 2010.3     \\

   $2\,^1S_0$      &  2581         & 2581        &  2554  &  0.84    & 2564(20)  \\

   $2\,^3S_1$      &  2643       &  2632       &   2642   &   0.84   & 2623(12)  \\
\hline
   $1\,^3P_0$      & 2399        & 2406         &  2366    & 0.73   & 2349(7)    \\

   $1P_1$          & 2456        &  2426        &    2420  &  0.73 & 2423(2)     \\

   $1P_1'$         & 2467       & 2469          &   2450   &  0.73  &  2427(51)       \\

   $1\,^3P_2$      &   2502      &  2460         &  2466   & 0.73   &  2465.4(1.3)  \\

   $2\,^3P_0$     &  2931        &    2919      &   2885   &  1.02   &          \\

   $2P_1$         & 2924          &  2932       &   2920   & 1.02  &       \\

   $2P_1' $       & 2961      &   3021           &  2969   & 1.02     & $D_J(3000)$  \\

   $2\,^3P_2$       & 2957       & 3012        &  2979     & 1.02   & $D_J^*(3000)$   \\

\hline

   $1\,^3D_1$     & 2817         &  2788          & 2743   & 0.92      &  \\

   $1D_2$       &  2816           & 2806          &  2755   & 0.92     &   \\

   $1D_2'$      &  2845         &  2850         &  2765    & 0.92    & 2737(15)  \\

   $1\,^3D_3$       &2833        & 2863           &  2763   & 0.92  &  2763.5(3.4)  \\

   $2\,^3D_1$      & 3231        &  3228          & 3209    & 1.20    &       \\

   $2D_2$         &  3212      &     3259     &     3209    &    1.20     &  \\

   $2D_2'$       &    3248       &  3307      & 3230  &    1.20     &   \\

   $2\,^3D_3$      & 3226         & 3335           & 3221  & 1.20    &     \\

\hline

  $1\,^3F_2$      &  3132   & 3090   & 3059       &  1.08    &    \\

  $1\,{}^3F_4$     & 3113  &   3187    &  3079        &  1.08    &          \\

  $2\,{}^3F_2$  & 3490    &         &     3422    &   1.37  &   \\

  $ 2\, {}^3F_4$   & 3466    &   3610     &    3430   &  1.37    &   \\
\hline
\end{tabular}
\end{table}

In the GE potential  $V_0(r)$,
\begin{equation}
 V_0(r)=\sigma r - \frac{4\alpha_{\rm V}(r)}{3 r},
\label{eq.10}
\end{equation}
the two-loop vector coupling $\alpha_{\rm V}$ is defined here without fitting parameters; namely, we use the QCD vector constant
$\Lambda_V=465(15)$~MeV,  determined via the conventional $\Lambda_{\overline{MS}}(n_f=3)=315(10)$~MeV, known from perturbative
 QCD \cite{31} and lattice QCD \cite{32}. For $\Lambda_{\rm V}=0.465$~GeV the asymptotic value
$\alpha_{\rm crit}=0.5712$ is close to  $\alpha_{\rm crit}=0.60$, used in Ref.~\cite{6}, while in Ref.~\cite{14} the  larger $\alpha_{\rm crit}=0.84$ is taken.
However, the details of  the $\alpha_V(r)$ behavior, which are different in considered models,  do not practically affect the masses of high
excitations because of the small contribution from the GE potential $V_{\rm GE}$, namely, $\la 10\%$. We give below the sets of the
parameters, used here and in  Refs.~\cite{6,14}:
\begin{eqnarray}
 \mbox{Ref.~\cite{6}}& m_{u,d} =220~{\rm MeV},&  m_s=419~{\rm MeV}, \quad m_c=1628~{\rm MeV},~ C_0=-253 ~{\rm MeV}
\nonumber \\
 \mbox{Ref.~\cite{15}}& m_{u,d} =330~{\rm MeV}, & m_s=500~{\rm MeV}, \quad m_c=1550~{\rm MeV},~ C_0=- 300~{\rm MeV}\
 \nonumber\\
  \mbox{present~paper}, & m_{u,d}=0,~\quad\quad\quad &  m_s=180~{\rm MeV}, \quad m_c=1435~{\rm MeV},~  C_0 =0.
\label{eq.11}
\end{eqnarray}

Although these parameters are very different, nevertheless, the calculated masses of the ground states agree with each other and with the
experimental data, with the exception of  $M(1\,^3P_0)$ and $M(1\,^3D_J)~(J=2,3)$,  which in Refs.~\cite{6,14} occur to be  $\sim (50-100)$ ~MeV larger
than in experiment and in our calculations (see Table~\ref{tab.01}). For the first excitations ($n_r=1$) the predicted masses, as a whole,  also agree with each other,
with exception of the $D_J^*(2\,^3P_J),~(J=1,2)$ states, where the difference is $\sim 50$~MeV; even a larger difference, $\sim 100$~MeV,
is found for the $D^*(n\,^3F_4)~(n=1,2)$ state.

For further analysis it is convenient to introduce the size of the string, $d(nL) = \langle r \rangle_{nL}$, which varies in  the range
(0.45 - 0.92)~fm for most low states with $n_r=0,1$ (see Table ~\ref{tab.01}).  Just in this region the linear behavior of the CP was proved in
lattice QCD \cite{33} and the field correlator method \cite{28,34}. The linear CP successfully describes the properties of the HQ and
the ground states of light mesons with $d\la 1.0$~fm \cite{16,22}. However, it gives too large masses for high states with $d > 1.2$ fm, and
just to reach agreement with experiment  the flattened CP was introduced \cite{21}. From Table~\ref{tab.01} one can see that the sizes of the
$2D$ and $nF$ charmed mesons are large enough and therefore their masses can be shifted (see later).

Recently in the region (3.0 - 3.2)~GeV three new structures were observed \cite{2,3,4} and the mass one of them,  $D_2^*(3000)$,
$M=(3214\pm 78)$~MeV ($J^P=2^+$), is larger than  $M(2\,^3P_2)\sim 3000$~MeV, predicted in Refs.~\cite{6,14} (see Table~\ref{tab.01}), and later we consider
this resonance as a candidate for the $3\,^3P_2$ state. For two other resonances, the unnatural parity $D_J(3000)^0$ and
$D_J^*(3000)$ of the natural parity the quantum numbers are not determined yet and for them different assignments were
suggested from the analysis of the  $D$-meson spectra and their strong decays. In particular, the $D_J(3000)^0$ resonance is supposed to be
the $3\,^1S_0$ state \cite{6,8}, or the $2P_1'$ state \cite{7,12,13,35}, while $D_J^*(3000)^0$ is tentatively assigned to possibly be   the $1\,^3F_4$
state \cite{6,35}, or the  $3\,^3S_1$ state \cite{8}, and $2\,^3P_0$  \cite{9,13}.

In our calculations, as well as in Ref.~\cite{6},  the mass $M(2P_1')=2959$~MeV occurs to be a bit smaller than that of the
$D_J(3000)$ (its experimental value $M(D_J(3000)^0)=3008(9)$~MeV), which quantum numbers  $J^P=1^+$ were assumed in Refs.~\cite{7,12,13,35}.
However, this resonance can also be identified with the $3\,{}^1S_0$ state, which mass for the linear CP is larger,  $M(3\,^1S_0)=3042$~MeV
(see Table ~\ref{tab.02}), but due to decay channels acquires a mass shift (see next section). Also the mass of the $2\,{}^3P_0$ state,
$M(2\,{}^3P_0)=2885$~MeV, is rather small to be a candidate for the $D_J^*(3000)$ resonance, while the calculated mass $M(2\,^3P_2)=2979$~MeV
appears to be more close to that of the $D_J^*(3000)^0$ structure (with  $M(\rm exp.)=3008$~ MeV), which quantum numbers corresponding to the   
$2\,^3P_2$ state cannot be excluded. The possibility that this resonance could be the candidate for the $3\,^3S_1$ state will be discussed in the next section.

In all three models the masses of the higher resonances ($n_r\geq 2$) refer to the region above 3.2 ~GeV (see Table~\ref{tab.02}),
with  exception of the $3S$ states, which masses, $M(3\,^1S_0)\simeq 3042$~MeV and $M(3\,^3S_1)\simeq 3096$~MeV, are smaller due to the stronger GE interaction in the
$S$-wave states. These  high excitations have two characteristic features -- large sizes,  $d \sim (1.3-1.6)$~fm,  and large excitation energies (above the ground states),
$E(nL)\ge (1.3-1.7)$~GeV. It is of interest that the values of  $d(nL)$ and $E(nL)$ (see Table ~\ref{tab.02}) appear to be close to  those in high charmonium states,
calculated with the same universal potential and  the $c$-quark mass (they are given for comparison in Table~\ref{tab.03}), namely, 
$d(nL)\sim (1.4-1.5)$~fm for $\psi(nS)$ and $\chi_{c0}(nP)$, if $n_r\geq 2$ \cite{16,23}. As seen from Table~\ref{tab.03}, for the
linear+GE potential the masses of high charmonium states exceed the experimental numbers, while in the flattened CP
a correlation between the size and the  mass shift of a given state is observed:

1. The states with $d\la 1.0$~fm  have no mass shifts (the large hadronic shift of $\psi(4040)$  has a different nature,
related to the strong coupling of $\psi(4040)$ to the nearby $D^*D^*$ threshold);

2. The states with  $d\simeq (1.1-1.2)$~fm have small shifts,  $\sim (20-30)$~MeV;

3. The states with large $d \geq 1.4$~fm have large shifts,  $\gtrsim 100$~MeV. In the case of $\chi_{c0}(4700)$, if it is interpreted as the $5\,^3P_0$ charmonium state,
the mass shift is obtained to be very large, $ \sim 200$~MeV.

Thus in charmonium  the picture is similar to that in light mesons, where the radial excitations with large sizes have also
large mass shifts and precisely the analysis of the mass shifts has allowed for the  extraction of  the parameters of the  flattened CP \cite{21}.

The hyperfine and fine-structure corrections to the centroid mass were calculated here, as in Ref. ~\cite{14}, and briefly presented
in the Appendix. In the RSH the corrections $\delta_{\rm hf}$ and $\delta_{\rm fs}$  are proportional to  the kinetic energies
$(\omega_q \omega_c)^{-1}$, but not to the constituent masses $(m_q m_c)^{-1}$ as it is in the potential models. This change follows from the general representation of
spin-dependent potentials in the field correlator method \cite{30}. Also in all  hyperfine corrections we take the universal strong coupling $\alpha_{\rm hf}=0.33$ \cite{36},
while the tensor and the spin-orbit splitting are determined by the same coupling $\alpha_{\rm V}(r)$, as in the GE potential  Eq.~(\ref{eq.10}).

\begin{table}
\caption{The masses (in MeV) and the sizes $d(nL)$ (in fm) of high $D$ meson for the linear+ GE potential with the parameters from (\ref{eq.11})}
\label{tab.02}
\begin{tabular}{|l|r|r|r|r|}\hline

State &     $ d(nL)$  & GM \cite{6} & EFG \cite{15}& this paper \\

\hline

  $3\,{}^1S_0$ & 1.20   & 3068   & 3062 & 3042   \\

 $3\,{}^3S_1$ & 1.20   & 3110    & 3096 & 3096     \\

 $4\,{}^1S_0$ & 1.42   & 3468   & 3452  &  3445   \\

 $4\,{}^3S_1$ & 1.42   & 3497   & 3482  &  3485  \\

 $5\,{}^1S_0$   & 1.63    & 3814   & 3793  & 3794     \\

 $5\,{}^3S_1$ & 1.63    & 3837   & 3822   & 3832   \\

\hline

 $3\,^3P_0$   &1.32     & 3343   & 3346  & 3304 \\

 $3\,^3P_2$   & 1.32    & 3353   & 3407   & 3339\\

 $4\,^3P_0$   & 1.55     & 3697  &        & 3657  \\

 $4\,^3P_2$   & 1.55     & 3701    &      & 3694    \\
 \hline

 $3\,{}^3D_1$ & 1.46   & 3588    &       & 3533     \\

 $3\,{}^3D_3$ & 1.46    & 3579    &       & 3601    \\

\hline
\end{tabular}
\end{table}

\begin{table}
\caption{The sizes  $d(nL)$ (in fm) and the masses of the $n\,{}^3L_J$ charmonium states  (in MeV) for the linear+ GE potential
(\ref{eq.11}) and  for the flattened CP with the parameters (\ref{eq.17}) ($m_q=0,~m_c=1.435$~GeV), and the mass shifts $\Delta(nJ)$ (in MeV)}
\label{tab.03}
\begin{tabular}{|l||c|c|r|r|r|}
\hline
State         &        & Linear CP          &   The mass shift  & Flattened CP    &     experiment       \\
         & $d(nL)$      & $M(n\,{}^3L_J)$ & $\Delta(nJ)$  & $M(n\,{}^3L_J)$ & Ref.~\cite{5}\\
         \hline
$1\,^3S_1$    &  0.35    & 3101         &  0             &  3101          & 3096.9    \\
$2\,^3S_1$     &  0.69      &  3687      & -5           &    3682         &     3686.1  \\
$3\,{}^3S_1$   & 0.95       &     4107   &  -21          &   4086          & 4039(1)  \\
$4\,^3S_1$     & 1.18       &   4459       &  -45       &   4414        &  4421(4)  \\
$5\,^3S_1$     & 1.38        &  4770     &    - 123       & 4647           & 4643 (8)  \\
\hline
$1\,^3P_2$     & 0.57        &  3542      &  0           & 3542        & 3556.2(1)   \\
$1\,^3P_1$    &  0.57         & 3505      &  0            & 3505        &   3510.7(10) \\
$1\,{}^3P_0$   & 0.57         &  3408      & 0             & 3408        & 3414.7(3) \\
$2\,^3P_2$      & 0.85       &  3964      & - 11          & 3953        & 3927.2(2.6)  \\
$2\,{}^3P_1$  &   0.85         &3927        & - 11           & 3916        & 3871.7(2) \\
$2\,^3P_0$    & 0.85        &  3879       &    -11        & 3868         & $3862^{+26}_{32}$ \\
$3\,^3P_2$    & 1.09         & 4346        &  -35          &  4311        &  \\
$3\,^3P_1$     &1.09          & 4312        &  -35         &   4277         &  $4278^{+8}_{-6}$ \\
$3\,{}^3P_0$   & 1.09        &  4259        &   -35       & 4224          &      \\
$4\,{}^3P_2$    & 1.30       & 4651         &  - 75       &  4576            &     \\
$ 4\,{}^3P_1$  &   1.30      &  4638        &  - 75       &   4563          &    \\
$4\,^3P_0$  &    1.30        & 4598       &   -75           &    4523           &  4506(11)   \\
$5\,^3P_2$   &   1.49        & 4946  &    - 178(20)          &    4768 (20)      &         \\
$5\,^3P_1$    &   1.49        &  4930    &  - 178(20)         & 4752 (20)     &      \\
$5\,^3P_0$   &  1.49        &  4904            & -178(20)      & 4726 (20)    &   4704(10)   \\
\hline
\end{tabular}
\end{table}

At this point it is of interest to look at the charmonium spectrum, where the high excitations have also large sizes,  which are only  $\sim 15\%$ larger than
those of the charmed mesons. In the linear CP, as seen from  Table~\ref{tab.03},  the high $c\bar c$ excitations exceed the experimental masses by 
$\sim (40-100)$ ~MeV
(and by $\sim 170$~MeV  for the $5\,{}^3P_0$ state). For the charmonium family in Table~\ref{tab.03} we give also the mass shifts and the masses of
 the $nS$ and the $nP$ states in the flattened potential, where a good agreement with the experimental data is obtained \cite{16,23}.

\section{The mass shifts of the charmed mesons in the flattened + GE potential}
\label{Sec.III}

The flattened CP $V_{\rm fl}(r)$ was introduced to describe high radial excitations  of light mesons \cite{21}, otherwise with the linear CP
the slopes of the radial RT \cite{22} exceed the experimental values by $\sim (40-50)\%$. Later  \cite{16} the same flattened  CP was applied to charmonium and 
the correct values of the mass shifts were obtained (see Table~\ref{tab.03}), thus allowing to consider this phenomenological potential as universal and 
to apply it for charmed mesons. The flattened CP  has the following form,
\be
V_{\rm fl}(r)= \sigma(r) r, \;{\rm with}\; \sigma(r)=\sigma_0 ( 1 - \gamma f(r)),
\label{eq.15}
\ee
where the function $f(r)$ is given by 
\be
 f(r)=\frac{\exp(\sqrt{\sigma_0}(r - R_0))}{B + \exp({\sqrt{\sigma_0}}(r - R_0))}.
\label{eq.16}
\ee
The function $f(r)$ is very small,  $\sim 10^{-3}$, at small distances and approaches 1.0 at large  $r\sim 3.0$~fm. The parameters of $V_{\rm fl}$
were fitted from the analysis of the radial RT's with different $J^{PC}$ in light mesons \cite{16,21}:
\be
\sigma_0=0.182~{\rm GeV}^2,~\gamma=0.40,~B=20,~R_0=6.0~{\rm GeV}^{-1}.
\label{eq.17}
\ee
This modified static potential $\tilde{V}_0 = V_{\rm fl}(r) + V_{\rm ge}(r)$, with the set of parameters
Eq.~(\ref{eq.17}) and the strong coupling as in Eq.~(\ref{eq.10}), was used to determine the charmonium spectrum (see Table~\ref{tab.03}),
where the sizes of high states are close to those of the charmed mesons. As seen from Table~\ref{tab.03}, the masses of the  $n\,^3S_1$ and $n\,^3P_0$ (with $n=4,5$)
occur to be in good agreement with the experiment. The flattened CP produces two effects, which are absent in the linear CP:

1. The mass shifts and the sizes $d(nL)$ increase for growing $n_r$~($n=n_r+1$) and for $\chi_{c0}(5P)$  its size $d(5P)=2.1$~fm;

2. The kinetic energies $\omega(nL)$ become almost constant for all high states. For example, for the $n\,{}^3S_1$ states with $n=3,4,5$ and
the $n\,{}^3P_0$ states with $n=2 - 5$ their kinetic energies are practically equal, $\omega(nL)\cong 1.72(1)$ GeV. We define the mass shifts 
of high states as,
\begin{equation}
 \Delta(nJ,S=1)=M_{\rm flat} (nJ) - M_{\rm lin}(nJ),
 \label{18}
\end{equation}
where  the masses $M_{\rm lin}(nL)$ and $M_{\rm flat}(nL)$ refer to the case  with the linear and the flattened CP, respectively. The shifts
$|\Delta|$ increase from a value 35 MeV for the $3\,^3P_0$ state up to 178 MeV for the $5\,^3P_0$ state.
The calculated masses of high charmonium states  turn out to be in rather good agreement with experiment and the resonances  $\chi_{c0}(4500)$ and
$\chi_{c0}(4700)$ can be identified as the  $ 4\,^3P_0$ and $5\,^3P_0$ charmonium states \cite{16}. This analysis allows to assume that the flattened  CP 
can be treated as the universal CP at large distances and  applied to charmed mesons. The calculated masses of high $D$ mesons are given in Table~\ref{tab.04}, 
while the masses of low states, given in Table~\ref{tab.01}, do not change.

\begin{table}
\caption{The kinetic quark energies $\omega_i(nL)$ (in GeV), the string size $d(nL)$ (in fm), the masses $M(nl)$, and the mass shifts $\Delta(nL)$ (in MeV)
of high charmed mesons ($n_r\geq 2$) in the flattened+GE-potential}
\label{tab.04}
\begin{tabular}{|c|c|c|c|c|r|}
\hline
State      & $\omega_q(nL)$ & $\omega_c(nL)$ &$d(nL)$(fm) & $M(nL)$ & $\Delta(nL)$ \\

\hline$2\,{}^3S_1$  & 0.53      & 1.62       & 0.86          & 2749      & 0     \\
$3\,^3S_1$   & 0.66        & 1.64      &   1.26         & 3062     & -34 \\
$3\,^1S_0$  &  0.66        & 1.64      & 1.26            &  3008    &   - 34 \\
$4\,^3S_1$   & 0.68        & 1.65       &  1.88          &  3356  &   -129 \\
$4\,^1S_0$  &  0.68        & 1.65       &   1.88         &  3316  &  -129 \\
$5\,^3S_1$   & 0.71        &  1.67       & 2.20          &  3592   &  -240(20)  \\
$5\,^1S_0$   & 0.71        &  1.67       &2.20            & 3554   & -240(20)  \\
\hline
$2\,{}^3P_2$   &  0.66    &  1.63       &1.08         &  2968         & -11\\
$2\,{}^3P_0$    & 0.66    &  1.63   &    1.08   &     2874    &   - 11 \\
$3\,^3P_2$  &  0.67        &  1.64    & 1.58            &   3264 & -75  \\
$3\,^3P_0$  & 0.67         & 1.64     &  1.58          &  3229   &  -75  \\
$4\,^3P_2$  & 0.68          & 1.65    &   1.85         &   3491   & -206\\
$4\,^3P_0$  & 0.68         & 1.65      & 1.85          &  3451  &  -206\\
$2\,{}^3D_1$  &  0.70   &     1.64   &    1.32  & 3175   &  -34  \\
$2\,{}^3D_3$  & 0.70   & 1.64         & 1.32     &  3187     &  -34 \\
$3\,{}^3D_1$  &0.71      & 1.65   &    1.84      &   3423        & -130 \\
$3\,{}^3D_3$   & 0.71   & 1.65  & 1.84           & 3491     &  -130 \\
\hline
\end{tabular}
\end{table}

From Table~\ref{tab.04} one can see that the $3S$ states have already shifted down by 34 MeV and the calculated mass
$M(3\,{}^1S_0)=3008$~MeV  coincides with the experimental value, $M(D_J(3000))=3008.1$~MeV \cite{3}. The states $2\,{}^3P_J$ have a small mass shift, $-11$ MeV,
while  the $3\,{}^3P_J$ states are shifted down  by 75 MeV and the predicted mass $M(3\,{}^3P_2)= 3264$~MeV is in agreement with that of the $D_2^*(3000)$, equal to
$3214\pm 78$~MeV \cite{2}. For the state  $3\,{}^3P_0$ we predict its mass equal to  3229 MeV, which is  117  MeV smaller than that from the GM analysis \cite{6}.
Note that the mass shifts increase for growing $n$ and reach $\sim 200$~MeV for the $4\,{}^3P_0$ state. From our analysis we expect similar large shifts  
for unobserved yet charmed mesons in the region $\ga 3.2$ GeV.

In Table~\ref{tab.04} we have not given the uncertainties in the mass values, which originate from irregular behavior of the second derivative of 
the potential $V_{\rm fl}(r)$ and are estimated to be $\sim (10-15)$ MeV. For that reason there exist also uncertainties in the values of the mixing
angle of the states with $J=L$ and $n_r\geq 2$, which masses are not given in Table~\ref{tab.04}.

\section{Conclusions}

The charmed mesons were studied with the use of the RSH, which is defined by the following fundamental parameters: the string tension, the current
quark masses, and the QCD constant $\Lambda_{\overline{MS}}(n_f=3)$. We take into account a flattening of the CP at large distances and consider
the flattened CP potential as a universal, which was tested in the analysis of the light meson spectra and in charmonium.
Due to the flattening effect the sizes of high states increase and their masses are shifted down, being
 $(35 -200)$ MeV smaller  than those obtained in other relativistic quark models.

For the as yet unobserved states the masses $M(2\,{}^3P_0)=2874$~MeV, $M(2\,{}^3P_2)= 2968$~MeV, $M(2P_1')=2959$~MeV
and $M(3\,{}^3S_1)=3062$~MeV are predicted, while the mass $ M(3\,{}^1S_0) =3008$~MeV  is obtained in agreement with the mass of the $D_J(3000)$ resonance.
The calculated mass $M(3\,{}^3P_2)=3264$~MeV occurs to be in agreement with the mass $3214(78)$~MeV  of the $D_2^*(3000)$ meson, 
which can be identified with the $3\,{}^3P_2$ charmed meson. 
 
Small mass shifts down are also obtained for the orbital excitations:  $M(1\,{}^3F_4)=3034$~MeV, $M(2\,{}^3D_1)=3175$~MeV,
and $M(2\,{}^3D_3)=3187$~MeV.

For higher $S$-wave states, the  calculated masses, $M(4\,{}^3S_1)=3356(12)$~MeV and $M(4\,{}^1S_0)=3316(12)$~MeV, occur to be $\sim 150$~MeV
smaller than the ones found in the GM model \cite{6}.

\begin{acknowledgments}

A.M.B. is grateful to Prof. Yu.A. Simonov for useful discussions.

\end{acknowledgments}

\appendix

\section{The fine-structure and the hyperfine splitting}

In the string picture \cite{30} the tensor splitting $t(nL)$,
\begin{equation}
 t(nL) = \frac{4}{3}
  \frac{\langle \frac{\alpha_{\rm V}(r)}{r^3} \rangle_{nL}}{\omega_q\omega_c},
\label{eq.A.1}
\end{equation}
and the spin-orbit splitting for $J=L+1,J=L-1$,
\be
a_{so}(nL) = \frac{1}{2\omega_q \omega_c} (3 t(nL) - \sigma \langle r^{-1} \rangle_{nL}),
\label{eq.A.2}
\ee
are expressed via the kinetic energies and the masses of the states with $J=l+1$ and $J=l-1$  are given by
\begin{eqnarray}
M(J=L+1, S=1) & =  & M_{\rm cog} +L\, a_{\rm so} - \frac{L}{2 (2L+3)} t; \\
\nonumber
M(J=L-1, S=1) & = & M_{\rm cog} -(L+1)\,a_{\rm so}- \frac{L+1}{2(2L-1)} t .
\label{eq.A.3}
\end{eqnarray}
For the flattened  CP the kinetic energies are given in Table~\ref{tab.04}, where one can see that their values practically
do  not  increase for high states.

The hyperfine splitting of the $S-$wave states is also expressed via the kinetic energies:
\begin{equation}
\delta_{\rm hf}(nS) = \frac{8 \alpha_{\rm hf} |R_{nS}(0)|^2}{9\omega_q(nS)\omega_c(nS)},
\label{eq.A.4}
\end{equation}

\end{document}